\documentclass[12pt]{article}
\usepackage{amsfonts}
\usepackage{color}
\usepackage[active]{srcltx}
\usepackage{pstricks}
\usepackage{pst-node}
\usepackage{graphicx}
\usepackage{float}
\usepackage{epsfig}
\usepackage{subfig}
\usepackage{epstopdf}
\begin{document}

\def\ba{\begin{eqnarray}}
\def\ea{\end{eqnarray}}
\def\w{\wedge}

\begin{titlepage}
\title{ \bf Holographic Superconductors with the General $RF^2$-type Couplings }
\author{ \"{O}zcan Sert\footnote{\tt osert@pau.edu.tr} \hskip 0.5cm  \\
 {\small Department of Mathematics, Faculty of Arts and Sciences, Pamukkale University}\\
 {\small 20017 Denizli, Turkey }
}

\maketitle

\begin{abstract}

\noindent
We explore the effects of the general  non-minimally coupled  $RF^2$-type couplings     on the holographic s-wave superconductors numerically in the Schwarzschild-AdS background. 
We calculate the condensation and conductivity of the model for the  coupling parameters    $a_1$ and $\beta$. 
 We obtain  
 that the bigger deviations  of the parameter $a_1$ from the minimal case lead to  the larger deviations of  the  gap frequency  from the  universal value $\omega_g/T_c \approx 8$.
 Moreover  the  smaller  $\beta $ and $a_1$  cause to  gradually  stronger and narrower  coherence peak.

\vskip 1cm



\end{abstract}
\end{titlepage}

\def\ba{\begin{eqnarray}}
\def\ea{\end{eqnarray}}
\def\w{\wedge}

\section{Introduction}
 The anti-de Sitter/conformal field theory (AdS/CFT) duality \cite{Maldacena} provides    a powerful relation between a  theory  of gravity in (d+1) dimensional AdS bulk space-time  and  a quantum field theory 
 at the boundary of the bulk.
  On the boundary, basic properties of a d-dimensional superconductor such as; critical temperature, phase transition,  can  be reproduced by the (d+1)-dimensional holographic dual model\cite{Hartnoll1,Hartnoll2}.
The model has a charged scalar field and an  electromagnetic field  as minimally coupled
in  an   AdS background. 
An interesting extension of the minimal holographic model is to consider  a non-minimal model which involves a term non-minimally coupled  Weyl tensor and Maxwell field  in the Lagrangian \cite{Wu}-\cite{Zhao}. The model with  the conformal invariant Weyl tensor    is  obtained from the special combination of  the general $RF^2$-type  model of this study: $a_3=\gamma/3$, $a_1=a_2=\gamma$ and the others are zero.
  Another special combination of these non-minimal terms  may be obtained from a five dimensional Gauss-Bonnet gravity  via dimensional reduction \cite{Myers}.
 Holographic
properties of the later  non-minimal model  are investigated in Schwarzschild-AdS \cite{Zhao2} and  in the charged black hole background \cite{Cai}.

The general $RF^2$-type non-minimal models with  vanishing scalar field and cosmological
 constant  were considered to find solutions to the problems such as   dark
 matter,  dark energy,  gravitational waves and    primordial magnetic fields in the universe \cite{drummond}-\cite{sert2}.
In the previous article  \cite{Sert}, we have considered the holographic superconductor with  arbitrary and perturbatively small coupling constants,   and obtained a relation for critical temperature and condensation as approximate analytical solution.
 It is interesting to find how the description of the holographic superconductor change when we consider numeric solutions  to this model.



In this study we investigate numeric properties of the non-minimal model with arbitrary coupling constants $a_i$ and compare them with approximate analytic results   in the probe limit.
 We  obtain the effects of the coupling parameters $a_i$ on the condensation of the scalar operators and conductivity.

\section{The Non-minimal Model with Arbitrary Coupling Constants} \label{model}

 We will give the model in terms of  exterior differential forms. We denote orthonormal co-frames $e^a$ and metric $g=\eta_{ab} e^a \otimes e^b$ where
$\eta_{ab}=\mbox{diag}(-1,1,1,1)$ is the Minkowski metric.
The Hodge star $*$  determines the orientation of space-time as   $*1 = e^0 \w e^1 \w  e^2 \w e^3$, where
$\wedge$  is the exterior product. 
The curvature
2-form is denoted  by $R^a{}_b$  and it is obtained from antisymmetric Levi-Civita connection 1-form as $R^a{}_b = d \omega^a{}_b + \omega^a{}_c\wedge \omega^c{}_b $.  The interior product is denoted by $\iota_a$
 which satisfying   $\iota_b e^a=\delta^a_b$, where $\delta^a_b$ is the Kronecker symbol. Then the 
 shorthand notations are used:  $\iota_aF =F_a$, $\iota_{b}\iota_{a} F =F_{ab}$, $
\iota_a {R^a}_b =R_b$, $ \iota_{b}\iota_{a} R^{ab}= R $

We take the following Lagrangian density 4-form  in differential form language:
 \begin{eqnarray} \label{lagrange}
  \mathcal{L} &=&\frac{1}{\kappa^2}\left[ R_{ab} \w *e^{ab}+ \Lambda *1
                   + (de^a + \omega^a{}_b \w e^b)\w \lambda_a \right] \nonumber \\
        & &-\frac{1}{2} F\w *F   -D\psi^{\dag}\wedge * D\psi -m^2\psi^{\dag}\psi*1  \nonumber \\
    & &  + 2a_1 F^{ab} R_{ab} \w *F + 2a_2 F^a\wedge R_a \w *F   + 2a_3 RF\w
    *F
      \nonumber \\  & &  + 2a_4 F^{ab} R_{ab} \w F + 2a_5 F^a\wedge R_a \w F   + 2a_6 RF\w
      F 
   \end{eqnarray}
where $\kappa$ is the gravitational coupling coefficient,   $a_i$, $i=1,2,3,4,5,6$ are  the non-minimal coupling coefficients and    $\Lambda$ is the cosmological constant. We will take  the cosmological constant   as $\Lambda=6/L^2$  in term of  AdS radius $L$.
 In the Lagrangian, $\lambda_a$ is Lagrange
multiplier constraining connection to be Levi-Civita.
 The complex
scalar field (hair) is denoted by $\psi$  and covariant  exterior derivative of it,   $ D\psi =
d\psi + i A \psi$.  The electromagnetic field 
is denoted by  $F=dA$, which is exterior derivative of  the electromagnetic potential 1-form
$A$.

This Lagrangian is equivalent to the Lagrangian in \cite{Sert} for $a_i= c_i/4$ and $a_4=a_5=a_6 = 0 $. If we set $a_1=a_3=\alpha $,\  \ $a_2=2 \alpha $ and the others are zero, we obtain the $RF^2$ corrected model which is investigated in  \cite{Zhao2}.

 In the probe limit, we have  the electromagnetic field equation and the
scalar field equation which is  obtained from independent variations of the non-minimal Lagrangian according to $A$ and $\psi^\dagger$, respectively,
\begin{eqnarray}
 d\Big\{ 4a_1*F^{ab}R_{ab}  + 2a_2\big[ R_a\wedge \imath^a*F-R*F+*(F^a\wedge R_a)\big]\label{Maxwellfe} 
  + 4a_3R*F -*F  & & \\ 4a_4( F^a\wedge R_a + F_{ab} R^{ab}) + (2a_4 - 2a_5 +4a_6) FR  \Big\}  - 2|\psi|^2*A = 0\  , &&\nonumber
    \end{eqnarray}
  \begin{eqnarray}
   D*D\psi -m^2\psi*1 &=& 0 \, . \label{scalarfe}
  \end{eqnarray}
It can be considered that $\psi$ is  real and  have the mass
$m^2=-2/L^2$; which is above the Breitenlohner-Freedman bound,
$m^2L^2 \geq -9/4$. We look for solutions    to the field  equations
in the background of   a planar Schwarzschild-AdS metric:
 \begin{equation}
    g = -f(r)dt^2  +  \frac{dr^2}{f(r)} + \frac{r^2}{L^2}(dx^2 + dy^2)
 \end{equation}
where
  \begin{equation}
            f(r) =\frac{r^2}{L^2}(1-\frac{r_H^3}{r^3}) \, .
   \end{equation}
We  calculate the field equations (\ref{Maxwellfe}) and (\ref{scalarfe}) for the metric, the scalar field and electric potential 1-form   which have only radial dependence  $\psi=\psi(r)$ and $A=\phi(r)dt$.  Then we change  the  variable $r$ to $z$  using this relation  $z=r_H/r$. Thus  the
outer region of the black hole  $r_H \leq r < \infty$  turns to the interval $0 < z \leq 1$ in this new coordinates.  

We look for solutions to  the following differential equations  which are obtained from  (\ref{Maxwellfe}) and
(\ref{scalarfe}):
 \begin{eqnarray}
    \left[1 + \frac{8a_1}{\beta L^2}(1-z^3)\right]\phi_{zz}  - \frac{24a_1}{\beta L^2}z^2\phi_z
    - \frac{2L^2}{\beta}\frac{ \psi^2}{ z^2(1-z^3)} \phi =  0 \, ,\label{phi1}\\
     \psi_{zz}- \frac{2+z^3}{z(1-z^3)}\psi_z + \left[\frac{L^4\phi^2}{r_H^2(1-z^3)^2}  +
         \frac{2}{z^2(1-z^3)}\right]\psi = 0 \, , \label{psi1}
 \end{eqnarray}
where $\beta= 1-24a_2/L^2 +48a_3/L^2$ and the subindex $z$ denotes
$d/dz$. We see that there is no contribution from the terms with $a_4, a_5, a_6$ for the electromagnetic potential $A=\phi(r)dt$.   We will take  $L=1$ and $r_H =1$, and focus first  on the different numerical values on the coupling constants  $\beta$ and $a_1$ to determine new properties of  the general  $RF^2$-type corrections. 

In the limit $z\rightarrow 0$,  the scalar field  $\psi$ and the electric potential function $\phi$ will have the following asymptotic behavior:

   \begin{eqnarray}
      \phi(z)=\mu - \rho z \, , \quad  \psi(z)=\psi_1z+\psi_2z^2 \, .
      \label{solasy}
   \end{eqnarray}
Here $\mu$ is the chemical potential  and $\rho$  is the charge density  on boundary.   $\psi_1$ and $\psi_2$ are related to vacuum expectation value of the condensation  and source operators in the field theory, that is  $<\mathcal{O}_1>  $ and  $<\mathcal{O}_2>  $. Since the mass ($m^2=-2$) is  near  the BF bound, we  choose   either $\psi_1=0$ or $\psi_2=0$ \cite{gubser}. 
We have also the regularity condition at the horizon $z=1$:

 \begin{eqnarray}  \label{sagasy}
     \phi(1)=0 \, , \quad  \psi_z(1)=\frac{2}{3}\psi(1) \, .
     \label{boundary}
   \end{eqnarray}

   We  obtain numerical solutions of the  general non-minimal $RF^2$-type holographic superconductors with the boundary conditions (\ref{solasy}) and (\ref{sagasy}).
   In figures 1 and 2, we give the solutions of the condensates  $<\mathcal{O}_1>  $ and  $<\mathcal{O}_2>  $      depending on the relative  temperature ${\frac{T}{T_c}}$ for different    $\beta$ and $a_1$. 
   The coupling constants $a_2$ and  $a_3$ 
   			appears as the  same combination  $ 1-24a_2 + 48a_3 = \beta$ in these  differential equations (\ref{phi1}), (\ref{psi1}) and (\ref{Ax}). Thus  one can choose either  $a_2 =0 $ or $a_3= 0 $ without loss of generality.
   		In figure 1, the cases  with  				
   			$\beta=2.44, 1.96, 1.48, 1.24, 1, 0.76, 0.52, 0.28$  
   			are obtained by setting    $a_2=0$ and
   			$a_3= 0.03, 0.02, 0.01, 0.005, 0, -0.005,- 0.01,  -0.015$, respectively.  These values are chosen  small for satisfying causality conditions. 
Similar to the previous findings on the holographic superconductors, the graphs  show that  the condensates $<\mathcal{O}_1> $ and $<\mathcal{O}_2>  $ vanish  as $T\rightarrow T_c$. Furthermore, the increase of    $\beta$ and $a_1$   makes the condensation gap larger similar to  the $RF^2$  corrections in \cite{Zhao2}.
To increase  $\beta $, \  $a_2$ must be decreased and/or $a_3$ must be increased. Thus, while the coupling terms $R_{ab} F^{ab} \wedge *F $ and $RF\wedge *F $  increase   the gap,  the term $R_a\wedge F^a \wedge * F$  decreases.

   \begin{figure}[H]{}
   	\centering
   	\subfloat[    ]{ \includegraphics[width=0.5\textwidth]{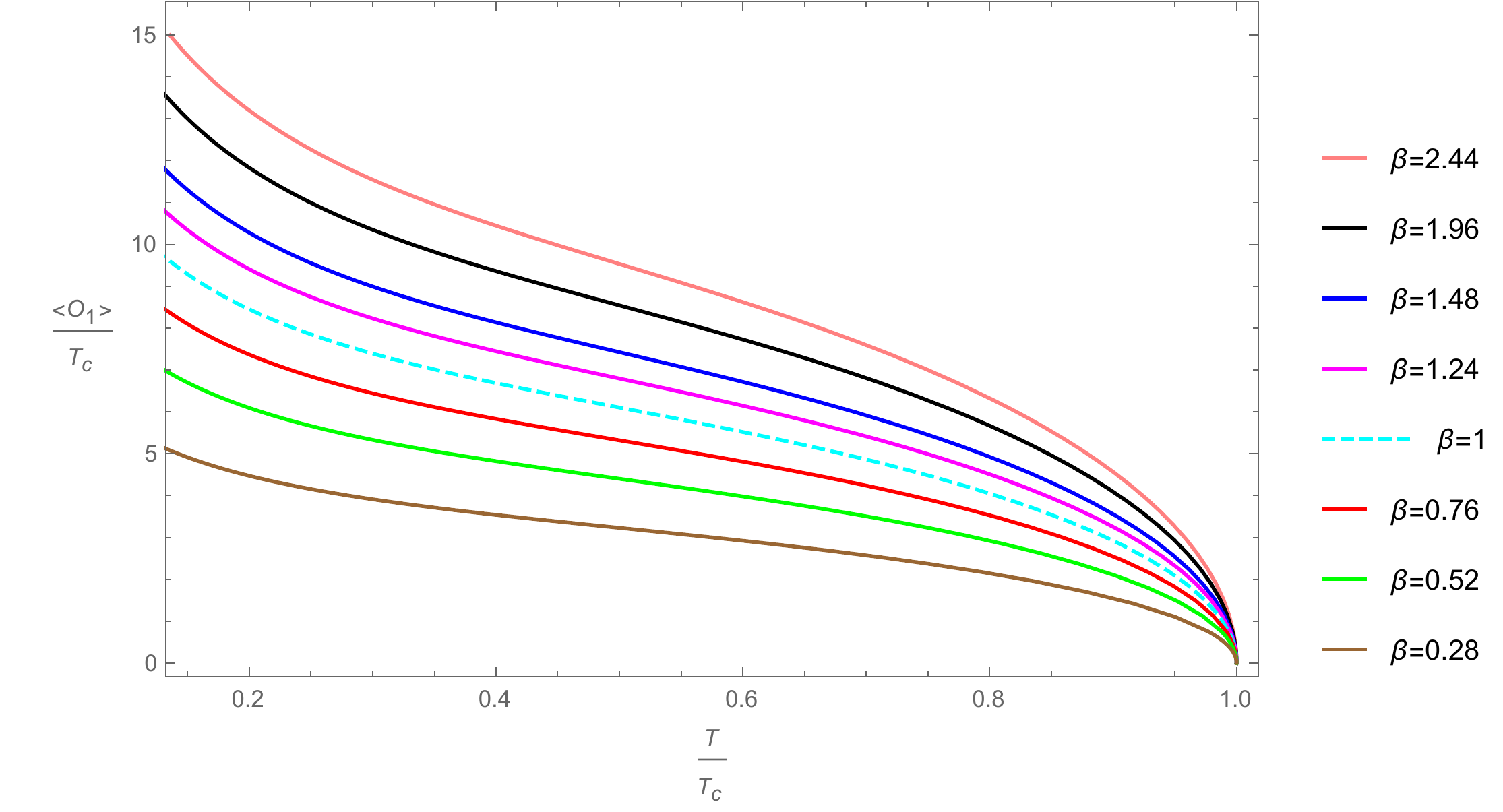} }
   	\subfloat[   ]{ \includegraphics[width=0.5\textwidth]{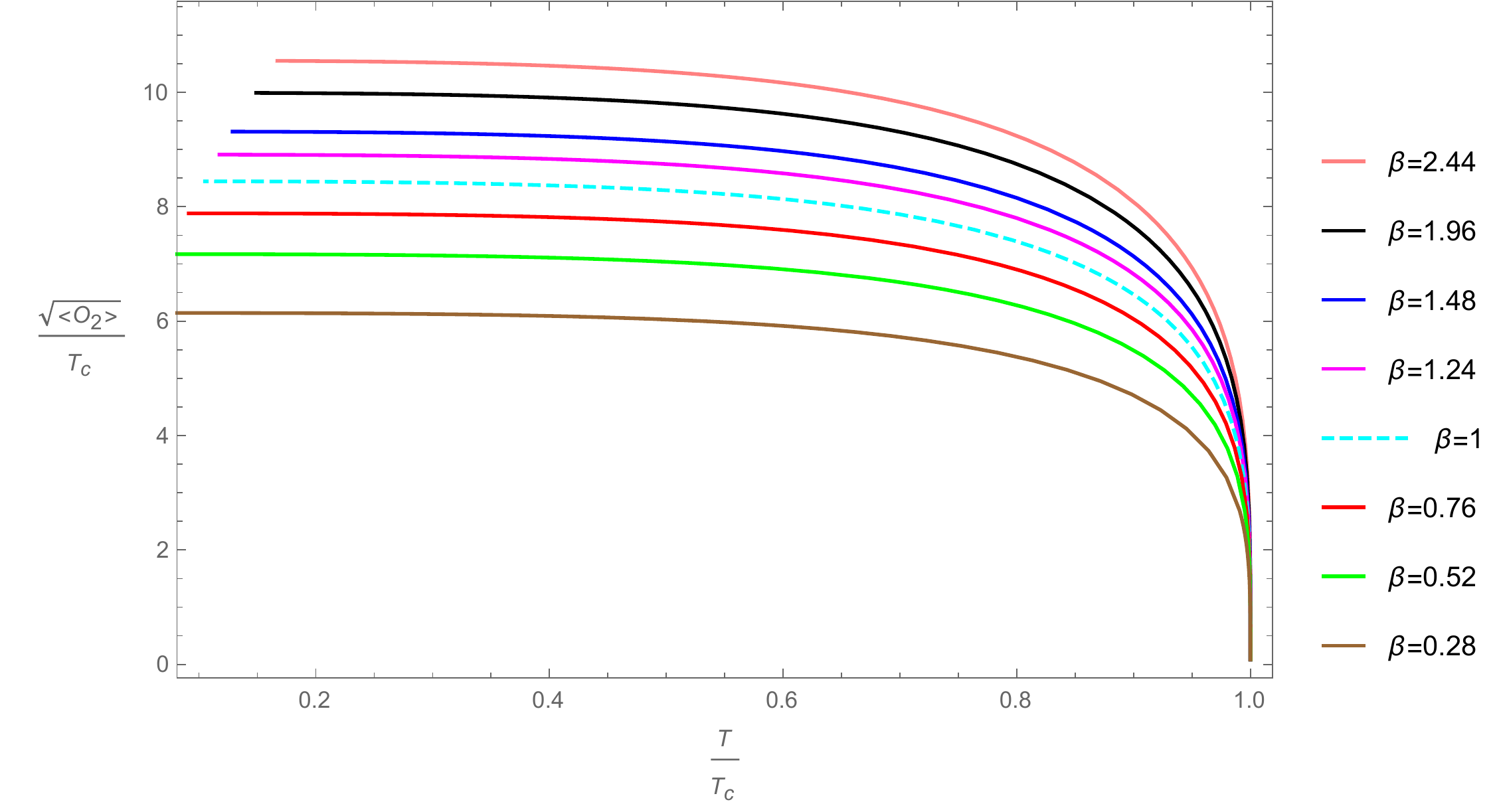}}  \\
   	\parbox{6in}{\caption{{{\small{     The condensates as a function of temperature:  (a) $<\mathcal{O}_1>$, \
                     (b)  $<\mathcal{O}_2> $, where $a_1=0$, $\beta=2.44, 1.96, 1.48, 1.24, 1, 0.76, 0.52, 0.28$ from top to bottom and dashed line is $\beta=1$.
   					}}}}}
   				\end{figure}   				                      
 
On the other hand,   we have   the the following approximate  analytical results which are  obtained  in  \cite{Sert}
   for  $<\mathcal{O}_2>$:
     \begin{eqnarray}
     \langle \mathcal{O}_2 \rangle = \frac{80\pi^2}{9}\sqrt{\frac{2\beta}{3}}TT_c
     \sqrt{1+\frac{T}{T_c}}\sqrt{1-\frac{T}{T_c}}\\
     \end{eqnarray}
     
          where the  critical temperature  is
          \begin{eqnarray}\label{Tc0}
          T_c=\frac{3\sqrt{\rho}}{4\pi L \alpha\sqrt{2\sqrt{7}}} \, 
          \end{eqnarray}
with $\alpha^2 = 1- 12a_1/(\beta L^2)$  .        
   As we see from these approximate analytic results,   $\beta$  has     an important effect   for increasing  the  $<\mathcal{O}_2>$  condensation gap, but   $a_1$ has  not,  which are consistent with the numerical solutions.
     \begin{figure}[H]{}
        \centering
        \subfloat[ $<\mathcal{O}_1>$   ]{ \includegraphics[width=0.5\textwidth]{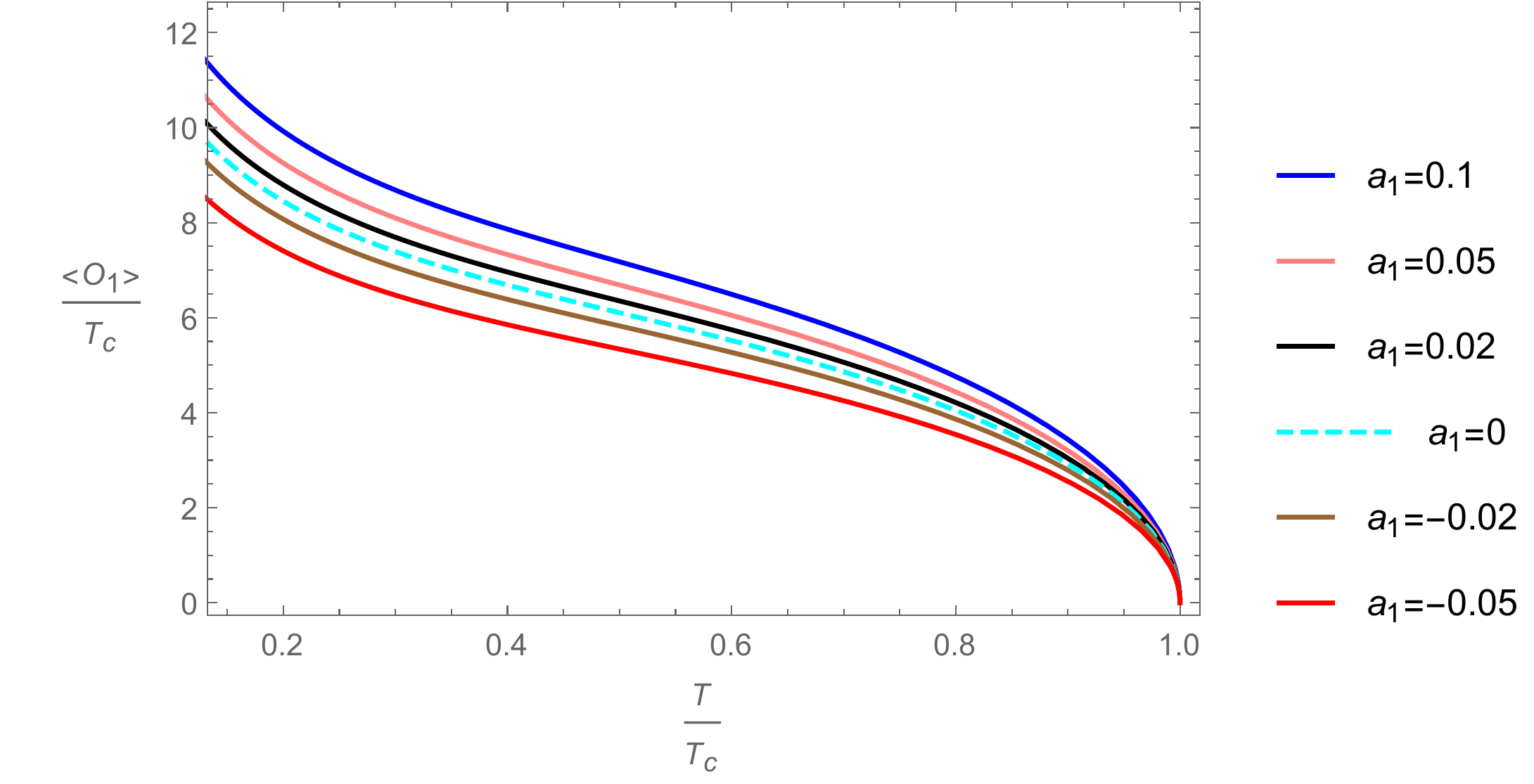} }
        \subfloat[ $<\mathcal{O}_2> $ ]{ \includegraphics[width=0.5\textwidth]{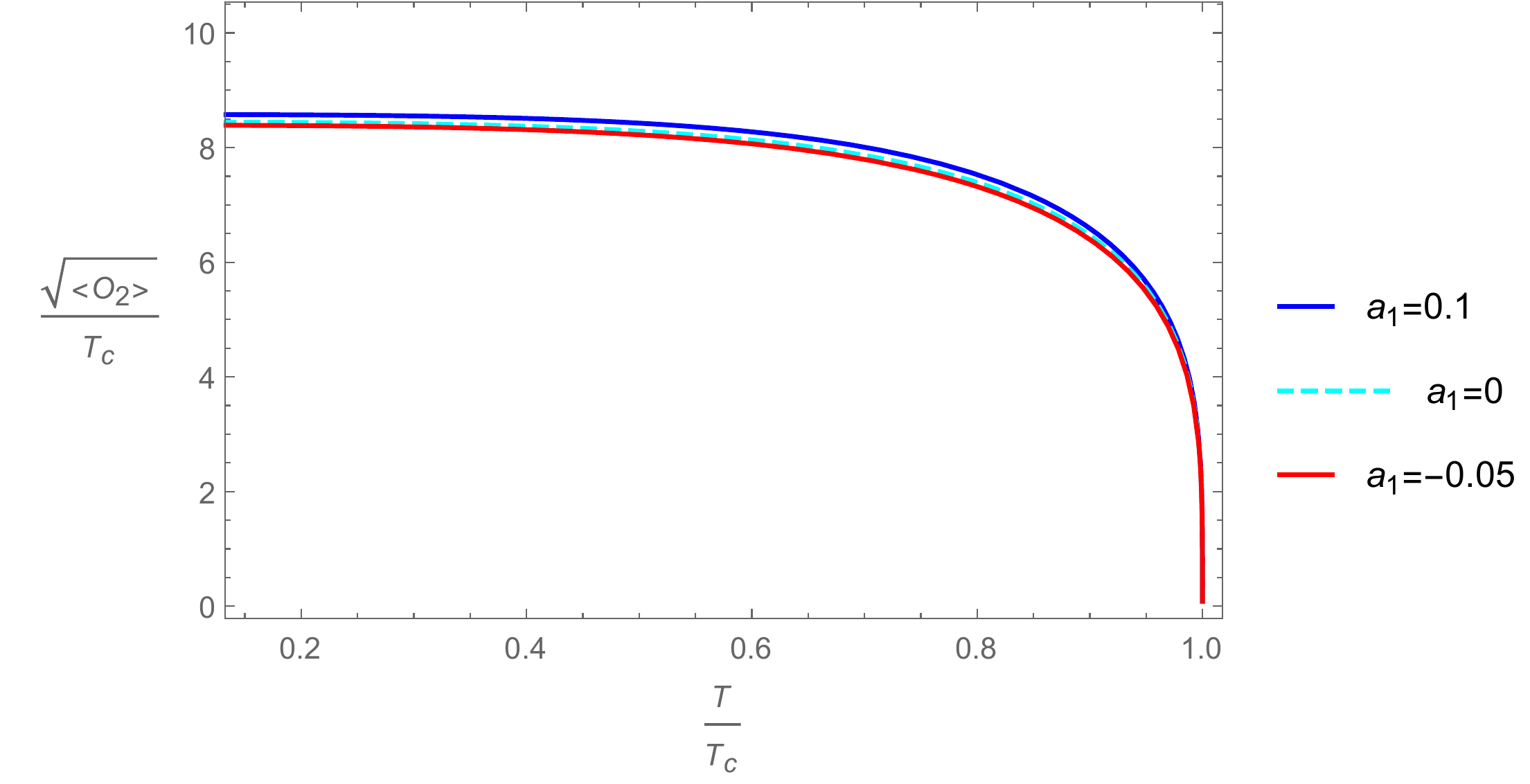}}  \\
        \parbox{6in}{\caption{{{\small{
                       The condensates as a function of temperature in unit $T_c$ for different  $a_1$:  (a) $<\mathcal{O}_1>$, \
                       (b)  $<\mathcal{O}_2> $, where $\beta=0$, $a_1=0.1, 0.05, 0.02, 0, -0.02, -0.05$ from top to bottom and dashed line is $a_1=0$.              
                     }}}}}
                  \end{figure}	
   The plots of  the second case   $a_1 \neq 0$ and $a_2=a_3=0$ (or $\beta=1$) can be seen in figure 2.   In this case,  $a_1$ has   less effects on  $<\mathcal{O}_2>$ than  $\beta$ while they  have more explicit  effects on  $<\mathcal{O}_1>$.
   We note that the condensates of  the general $RF^2$-type model have similar behaviors with the condensates of  the special combination of  $RF^2$ corrections in \cite{Zhao2}.
     But these results are   completely different from the effects of the Weyl corrections discussed in \cite{Wu}.

  \begin{figure}[H]{}\label{Tcg}
     \centering
     \subfloat[ $T_c-a_1 $  graph for $\beta= 0$ ]{ \includegraphics[width=0.5\textwidth]{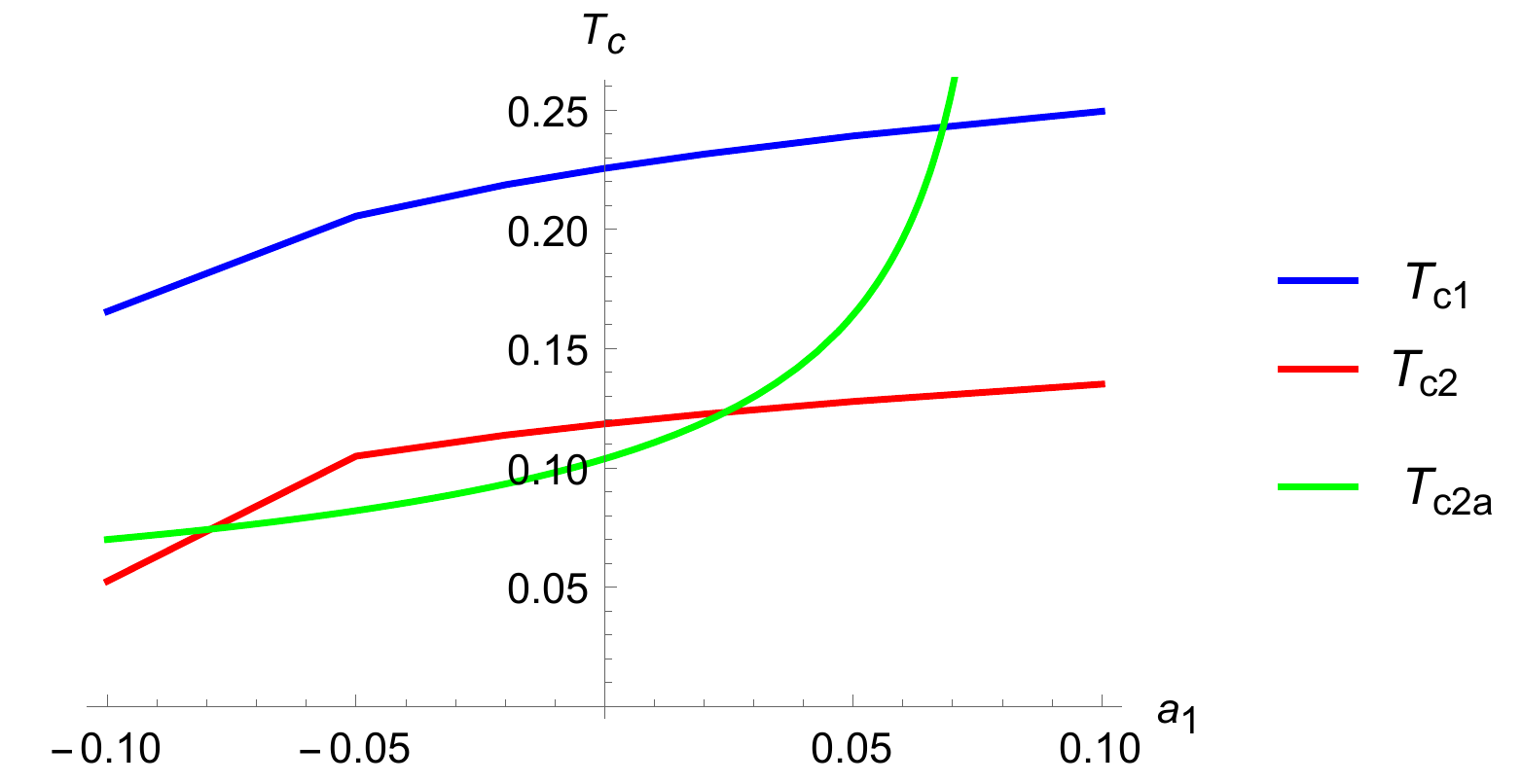} }					\subfloat[ $T_c-\beta $  graph for $a_1= 0$ ]{ \includegraphics[width=0.5\textwidth]{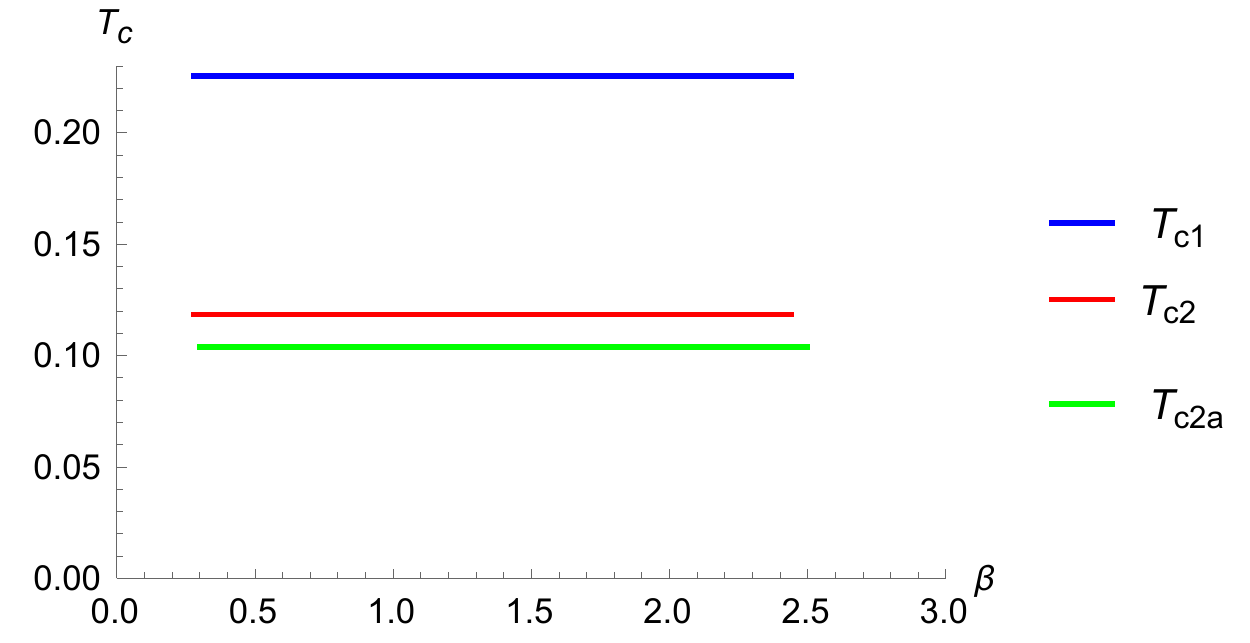}}\\
     \parbox{6in}{\caption{{{\small{ The numerical   graphs of the critical temperature     for the operators  $<\mathcal{O}_1>$ ($T_{c1}$),  \ \ $<\mathcal{O}_2>\ (T_{c2})$, and the approximate analytical graph for  $<\mathcal{O}_2> \ (T_{ca})$.
                  }}}}}
               \end{figure}
      				   				   				
    A comparison of 	the  numerical and approximate  analytical results for  the dependence of the critical temperature $T_c$ to the coupling parameters $\beta$ and $a_1$ can be found in figure 3.      The approximate analytic results  deviate from the numeric solutions after $a_1 = 0.05 $, excessively. This implies that there is an upper limit as $a_1< 1/12$ in the approximate analytic results.
     We see  from figure 3-(a) the critical temperature $T_c$ will increase as  the coupling parameter $a_1$  increases for the condensates. Thus the formation of scalar hair becomes more easier.  But $\beta$  does not   have  any important effect on the critical temperature for $a_1= 0$, see figure 3-(b). However,  when $a_1 \neq 0$, this situation  of $\beta$   will change.     We give    the formation of condensates    in figure 4 for the last case.

    \begin{figure}[H]{}
       \centering
       \subfloat[ $<\mathcal{O}_1>$   ]{ \includegraphics[width=0.5\textwidth]{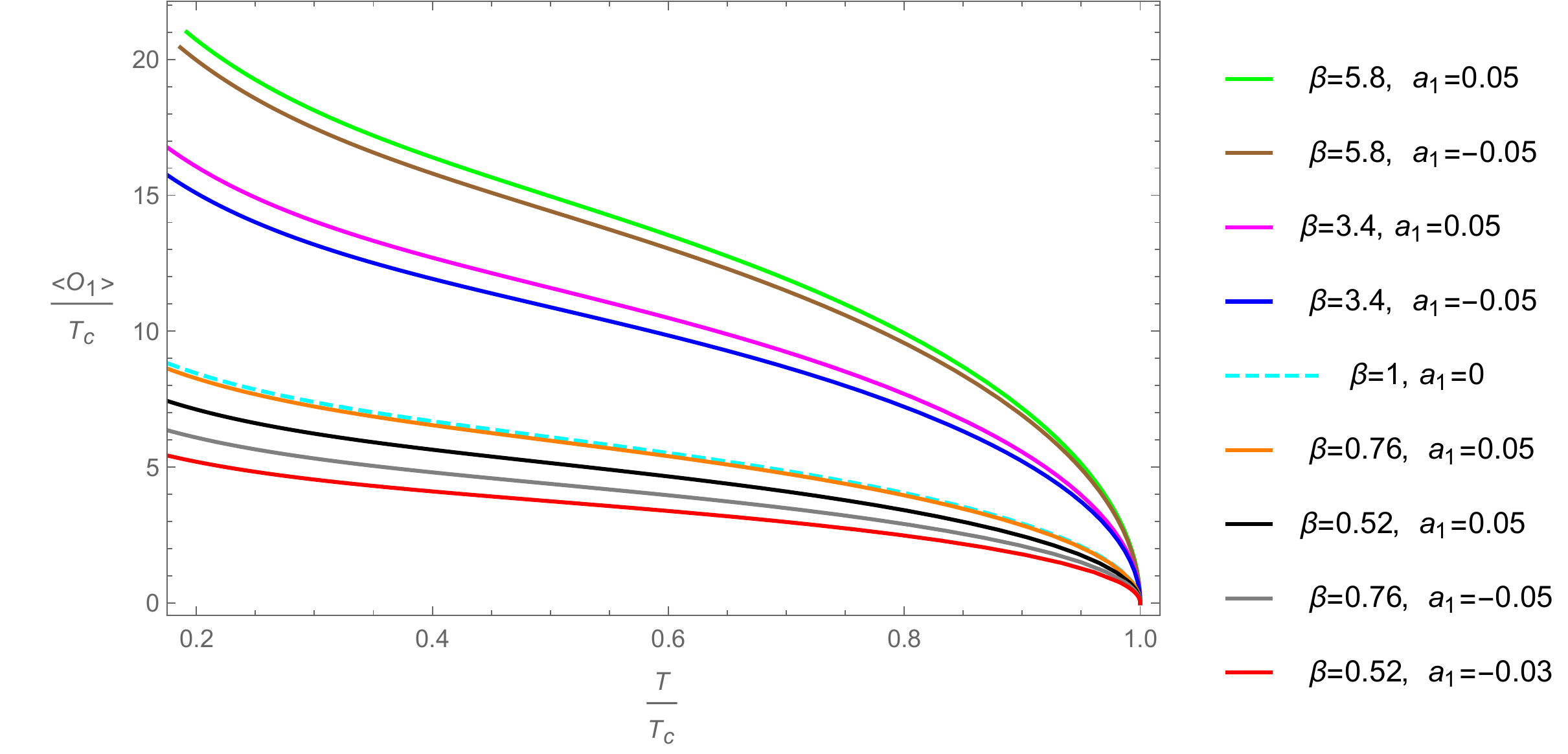} }
       \subfloat[ $<\mathcal{O}_2> $ ]{ \includegraphics[width=0.5\textwidth]{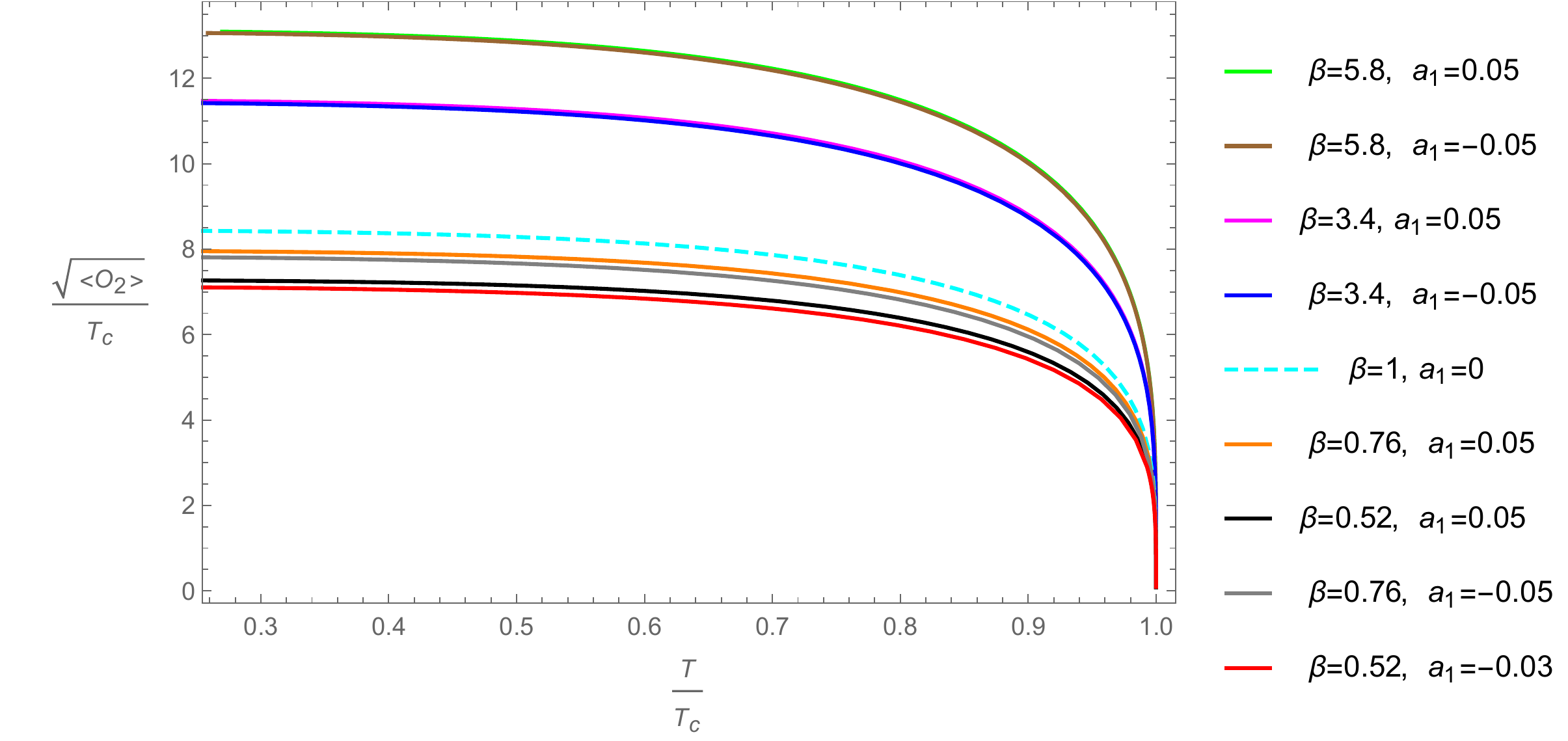}}  \\
       \parbox{6in}{\caption{{{\small{ The condensates as a function of temperature for $<\mathcal{O}_1>$ and \
                     $<\mathcal{O}_2> $.
             The condensation gap decrease as the parameters decreases from top to bottom.      }}}}}
                  \end{figure}

\section{Conductivity}
In this section, we investigate electrical conductivity of the general non-minimal $RF^2$-type  holographic superconductors.
Since the current one-form $J$ on the boundary is related with the Maxwell potential one-form in the bulk, we consider 
the perturbed Maxwell potential $\delta A_x = A_x(r) e^{-i\omega t} dx $, which has only x-component and find the following differential  equation from Maxwell field equation:

\begin{eqnarray}\label{Ax}
&& \left[ \beta + 4a_1(z^3 +2) \right] A_x''+ \left[ \frac{3\beta z^2}{z^3-1}  + \frac{12a_1z^2(2z^3+1) }{z^3-1 }\right] A_x' \nonumber \\
&&+ \left[ \frac{\beta \omega^2 }{(z^3-1)^2}   + \frac{4a_1 \omega^2(z^3 +2 )}{(z^3-1)^2} + \frac{2\psi^2}{z^3 -1 }\right] A_x =0
\end{eqnarray}

We have the following  ingoing wave  boundary condition  near horizon (z=1):
\begin{eqnarray}
A_x(r) \sim f(r)^{-\frac{i\omega}{3r_+}}
\end{eqnarray}
and the asymptotic AdS boundary condition at $z\rightarrow 0$:
\begin{eqnarray}
A_x = A^{(0)} +\frac{A^{(1)}}{r}.
\end{eqnarray}
The conductivity of the dual superconductor is calculated from 
\begin{eqnarray}
\sigma = - \frac{iA^{(1)}}{\omega A^{(0)}}
\end{eqnarray}

	We calculate numerically the conductivities depending on the frequency   $\frac{\omega}{T_c}$  using (\ref{Ax}), (\ref{psi1}), (\ref{phi1})  with the above boundary conditions. These results from the condensate $<\mathcal{O}_2> $ are illustrated in figure 5   for different $\beta$  values with $a_1=0$ and for different  $a_1$ values with $\beta =1$.
   	\begin{figure}[H]{}
            \centering
            \subfloat[ ]{ \includegraphics[width=0.5\textwidth]{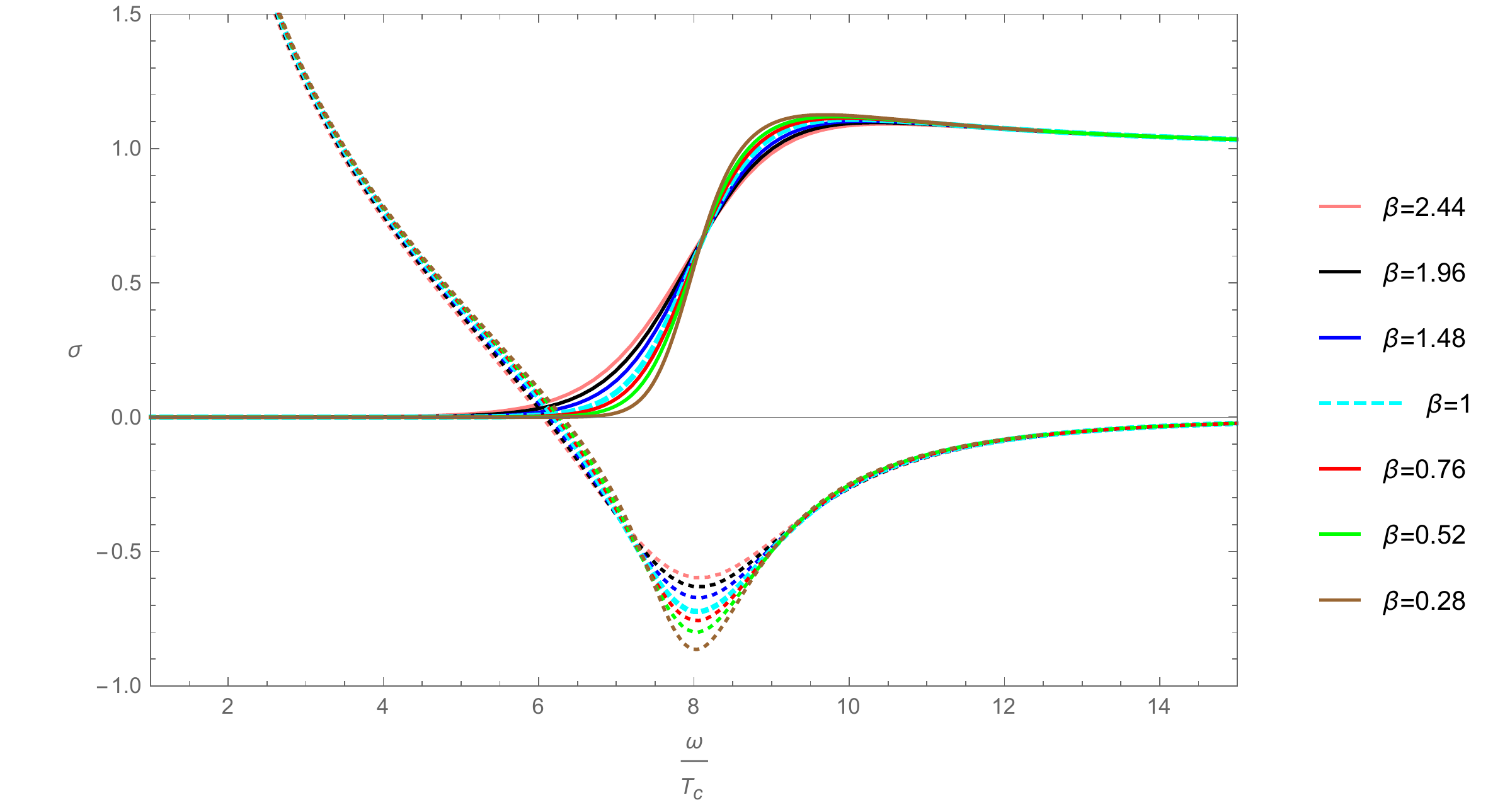}  } 
            \subfloat[    ]{   \includegraphics[width=0.5\textwidth]{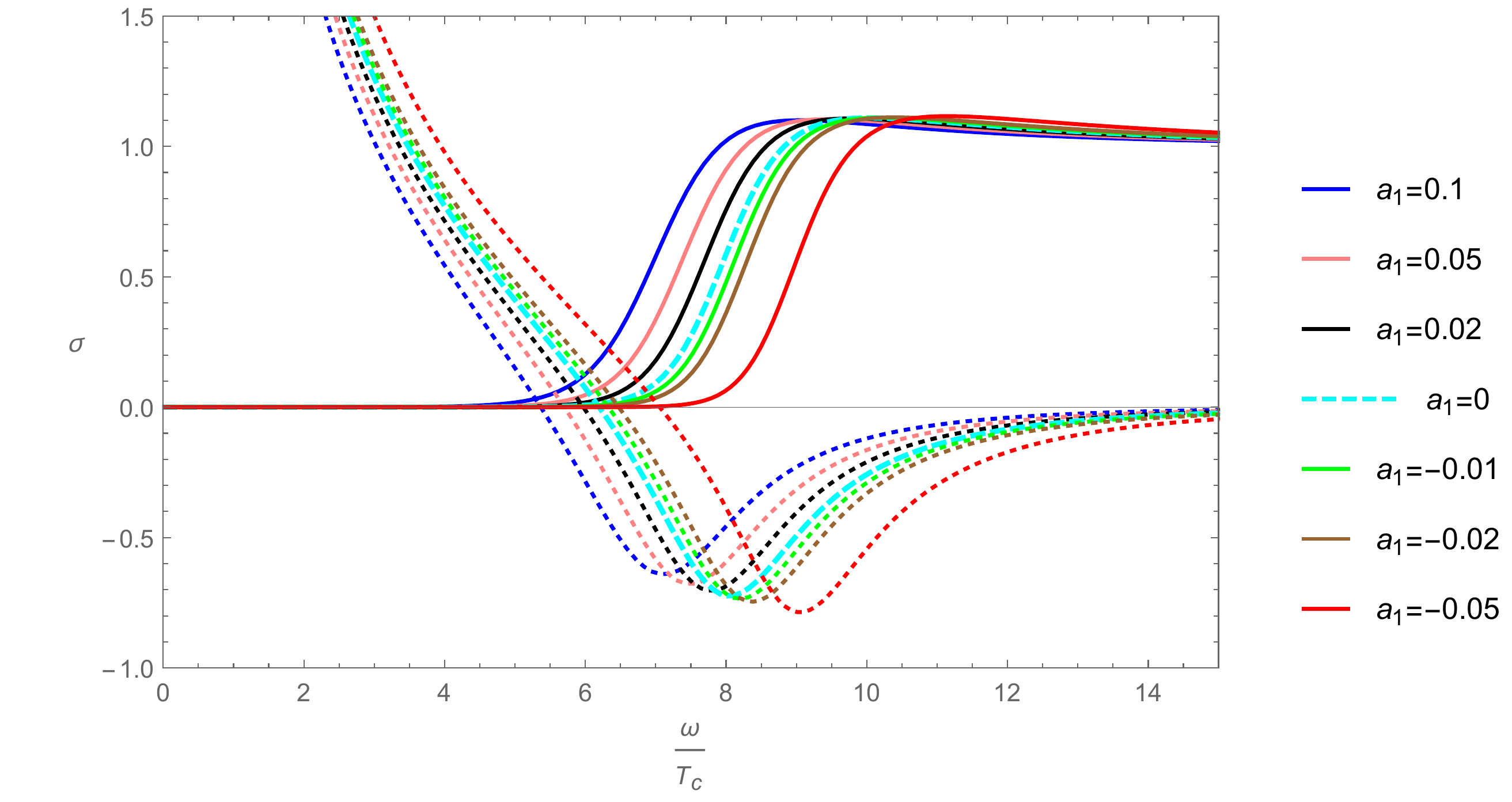}   }
         \\
            \parbox{6in}{\caption{{{\small{The conductivity as  a function of frequency in unit of critical temperature  for  $<\mathcal{O}_2> $.  While the solid lines represent the real part,  the dotted lines  represent  the imaginary part of conductivity,  and the dashed line (blue) is the case without non-minimal couplings ($a_1=0$ and $\beta=1$)  for $T/T_c \approx 0.2 $ .
                        }}}}} 
                     \end{figure}
In figure 5, we see that
there is a gap in the conductivity with the gap frequency  $\omega_g/T_c$  and the  gap     rises quickly near the gap frequency which  corresponds to the minimum of the imaginary part.   Furthermore,
 the value of the gap frequency decreases (from right to left)  as the parameter $a_1  $ increase (from bottom to top). But, while $a_1=0$,  $\beta$  does not have any important effect on  the value of the gap frequency and it has nearly the same  gap frequency with the  minimal coupling case ($a_1 = 0 $ and  $\beta=1$), that is;  $\omega_g/T_c \approx 8$. 
 On the other hand,   the effect of  the  smaller  $\beta $ or  $a_1$  is shown as   gradually  stronger and narrower  the coherence peak.


These results  support the previous findings  obtained  from the holographic models with the special  $RF^2$ correction in \cite{Zhao2} and Weyl correction in \cite{Wu}.
 This   shows that
the   correction terms with $a_1$, will have important effects  on
 the  gap frequency for the  strongly coupled model.

As we see the curves for $<\mathcal{O}_2>$ in figure 6,   there is another type of choice with $a_1\neq 0$ and $\beta\neq 1$ in the conductivity. In this case   the gap frequency 
has more  shifts  according to the case $a_1\neq 0$ and $\beta =  1$.
 The smaller  $\beta$ parameters   
  lead to the bigger deviations for $a_1 \neq 0$, especially; for $\beta< 1$.

 
    \begin{figure}[H]{}
                  \centering
                  \subfloat[ ]{ \includegraphics[width=0.5\textwidth]{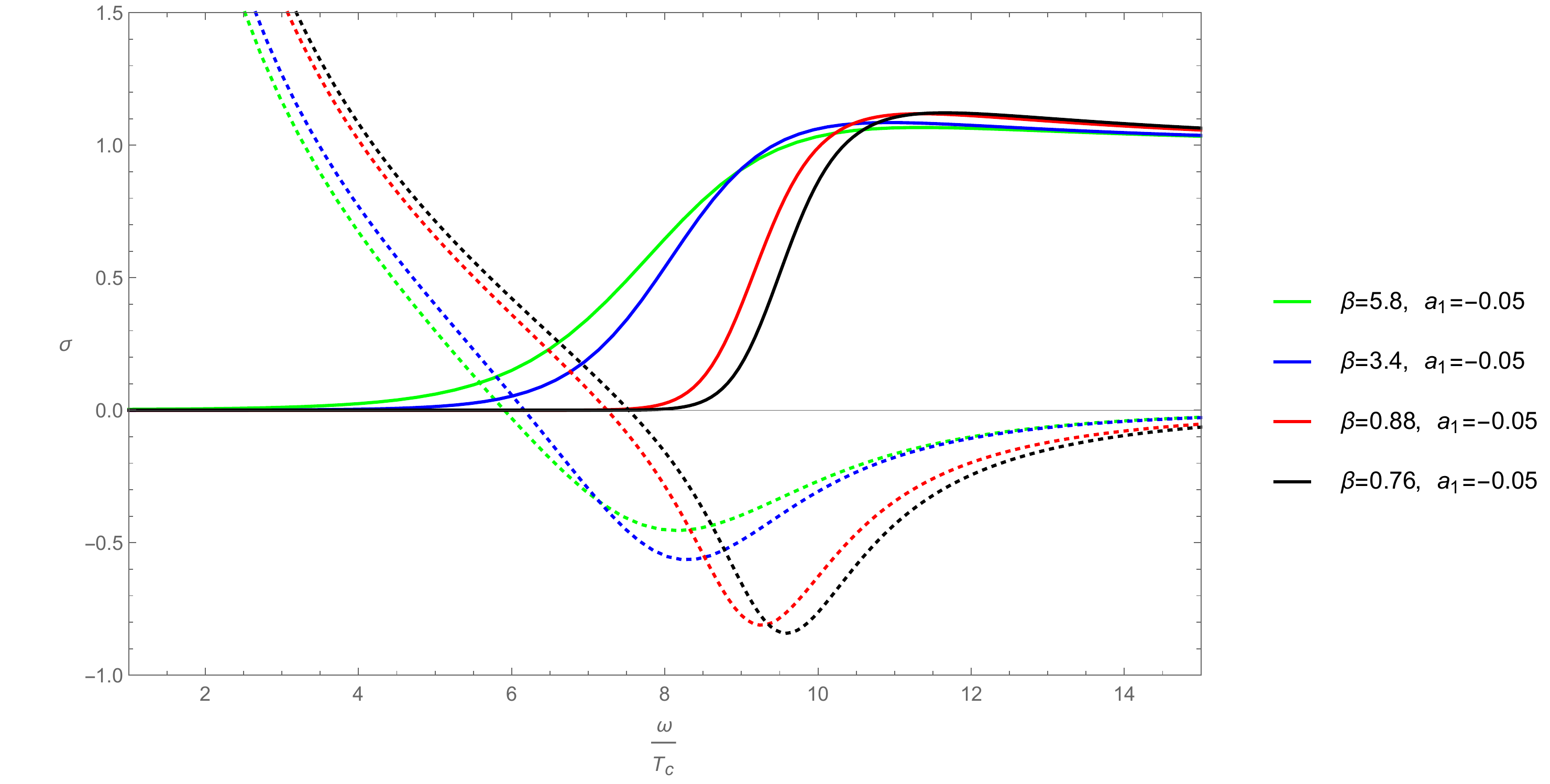} }
                     \subfloat[   ]{ \includegraphics[width=0.5\textwidth]{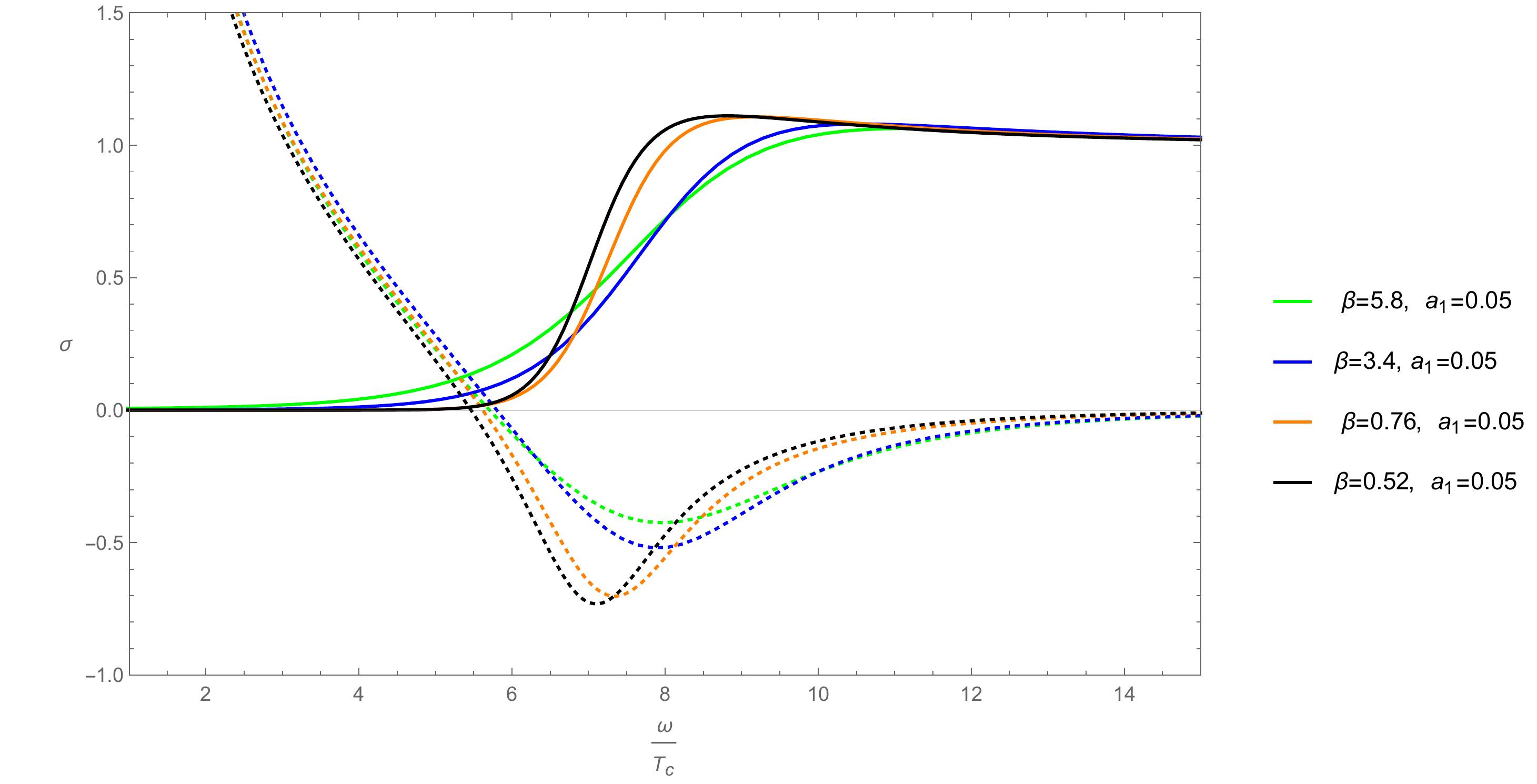}} 
                    \\
       \parbox{6in}{\caption{{{\small{ The conductivity which corresponds to the operator  $<\mathcal{O}_2> $ as  a function of frequency in unit of critical temperature  $a_1\neq 0$ and $\beta \neq 1$. 
                              }}}}}
                           \end{figure}
                           
     We have also  plotted  in  figure 7   the real part of  conductivity depends on the frequency which is  in  unit of the condensate. The curves from left to  right which corresponds to the conductivity for decreasing beta values from top to bottom.
     
     \begin{figure}[H]{}
        \centering
        \subfloat[     ]{ \includegraphics[width=0.5\textwidth]{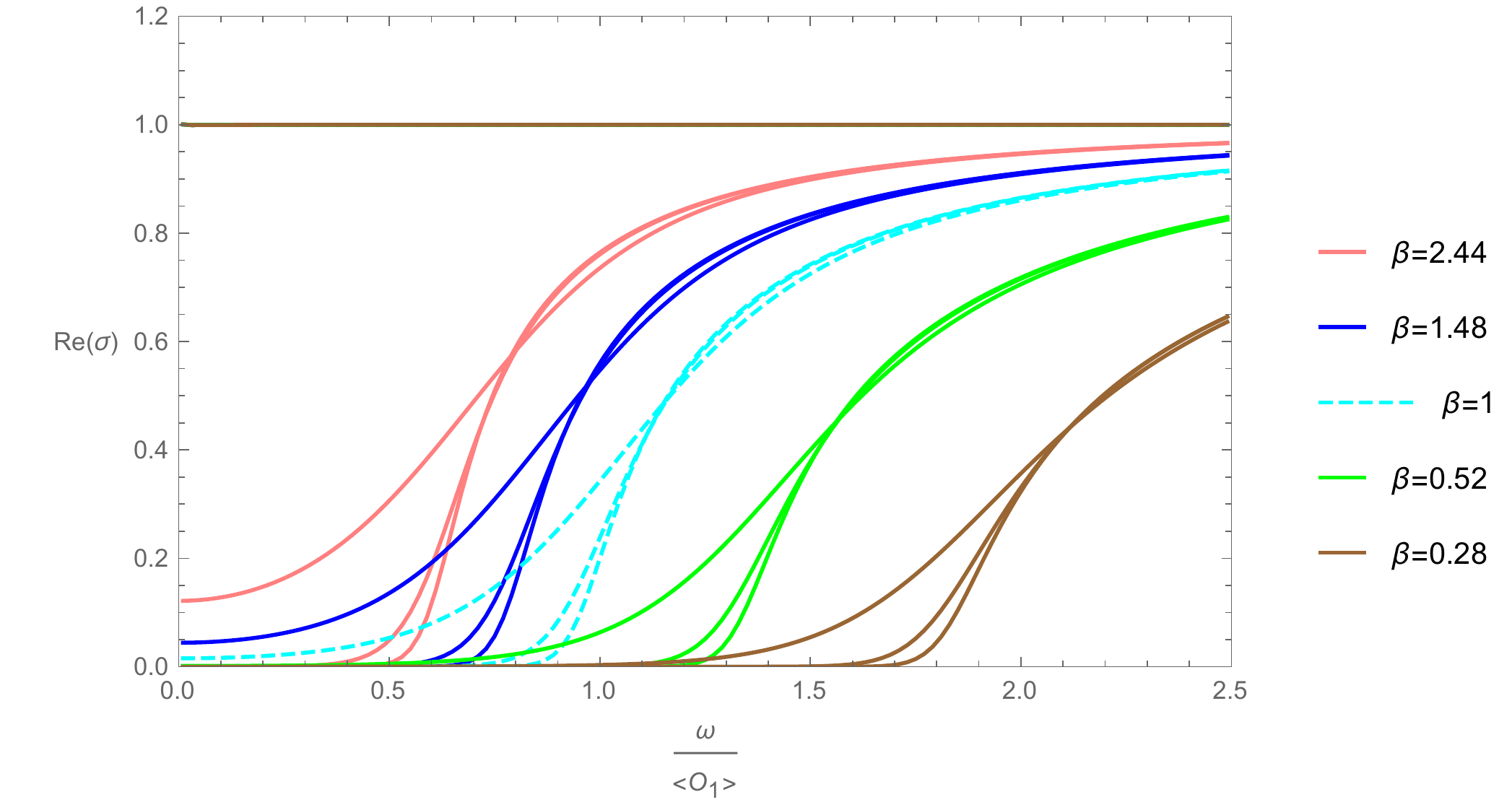} }
        \subfloat[   ]{ \includegraphics[width=0.5\textwidth]{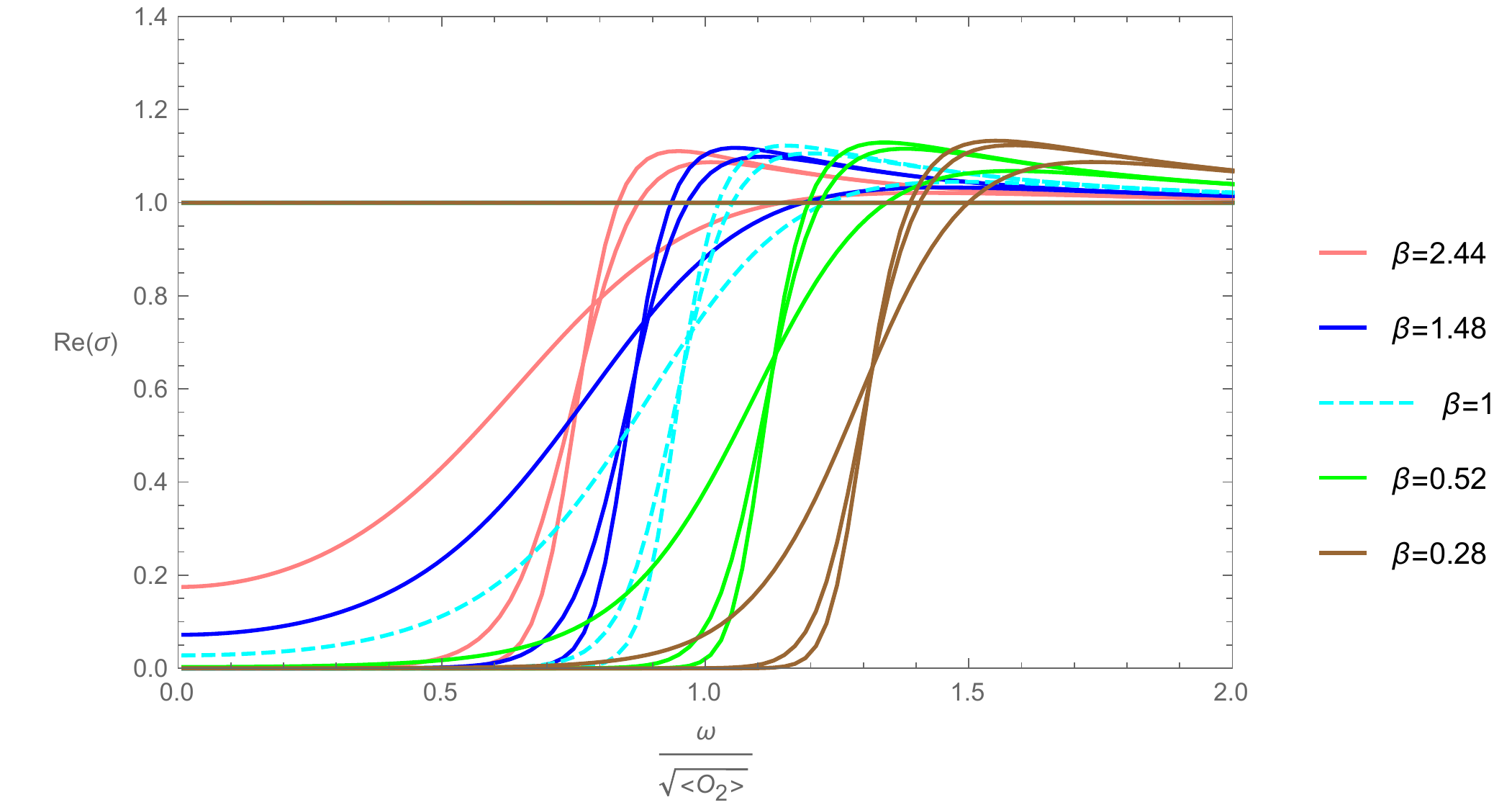}}  \\
        \parbox{6in}{\caption{{{\small{ The real part of conductivity as a function of frequency with  $a_1 = 0$ and $\beta \neq 0$  for \   $<\mathcal{O}_1>$\  (a)  and 
        					\ $<\mathcal{O}_2>$ \ (b). 
                       }}}}} 			
                  \end{figure}

  In the table below, we give  $T/T_c$ values of the curves in figure 7  from left to  right, respectively.

 \begin{table}[H]
 	  {\small 
      \centering
   \begin{tabular}{|l|c|c|c|c|c|}
               	\hline
      \ \	$\beta $   & 2.44 & 1.48 & 1 & 0.52 &0.28\\
                       		\hline
       $({\frac{T}{T_{c}})_a} $    & 0.67, 0.31, 0.20 & 0.57, 0.26,  0.16  &  0.50,  0.22, 0.14 &  0.39, 0.17, 0.11 &0.31, 0.14, 0.09\\
                       		\hline
      $({\frac{T}{T_{c}})_b} $ & 0.81, 0.36, 0.19 & 0.72, 0.27, 0.15 & 0.64, 0.22, 0.12 & 0.49, 0.16, 0.08 & 0.35, 0.11, 0.07\\
                       		\hline
     	\end{tabular}
                       	\caption{ $\frac{T}{T_c}$ values,  $(\frac{T}{T_c})_a$ corresponds to the approximate  values    in figure 7(a)    and  $(\frac{T}{T_c})_b$ in figure 7(b)    from left to right for each $\beta$, respectively. }
                       	\label{tab:template}
                       }
                       \end{table}
                                
From this figure,  we see that the horizontal lines in  the real part of the conductivity represent  the cases with  the temperatures $T\geq T_c$  and  in these cases  we can not observe any condensation.  This corresponds to the normal phase of the AdS boundary while  the conductivity has a gap for $T< T_c $ which is superconducting phase.  
 The gap  in the conductivity and   the coherence peak    shifted from right to left     with the increase of  $\beta$.	We have also found similar behaviors  in the conductivity for changing   $a_1$  with $\beta=1$.


\section{Concluding remarks}

We have investigated 
the  general non-minimal $RF^2$-type holographic superconductors in the probe limit. We have found that the increase of the coupling parameter $a_1$  leads to the higher critical temperatures while $\beta=1$ ($a_2=a_3=0$). But, $\beta$ which is the combination of coupling parameters $a_2$ and $a_3$  does not give any contribution    for higher critical temperature while $a_1 =0$. There is another choice  with $a_1\neq 0$ and $\beta\neq 1$. In this case, $\beta $ also have important effects for higher critical temperatures. 
We have shown that the higher correction terms with $a_1$ and $\beta $  lead to the larger    condensates $<\mathcal{O}_1>$ and $<\mathcal{O}_2>$. 
These results are  consistent with  the findings for $<\mathcal{O}_2>$  in the  approximate analytical calculations \cite{Sert}.

 We also calculated the conductivity of the general $RF^2$-type holographic superconductors.
We found that the parameter $a_1  $  has more effect on  the gap frequency than $\beta$.
Furthermore,  $\beta $ and $a_1$  change the strength and amount of the conductivity and coherence peak.

 In figure 6, we  saw       the another  type of choice with $a_1\neq 0$ and $\beta\neq 1$, and found     the gap frequency
 has more  shifts   from  the case with only $a_1\neq 0, \beta =1 $ or $\beta \neq 1, \ a_1=0$.
 The smaller  $\beta$ parameters   
 lead to the bigger deviations for the values $a_1 \neq  0 $.
 These results  support the previous findings  obtained  from the holographic models with the special  $RF^2$ correction in \cite{Zhao2} and Weyl correction in \cite{Wu}.
 This   shows that
the   non-minimal corrections change 
the universal  gap frequency.

 \section*{Acknowledgement}
 
 \noindent This work is supported by the
 Scientific Research Project (BAP) 2016HZDP023, Pamukkale
 University, Denizli, Turkey.


\begin{thebibliography}{99}

\bibitem{Maldacena} J. Maldacena, Adv. Theor. Math. Phys. 2, 231 (1998) [Int. J. Theor. Phys. 38, 1113 (1999)]

\bibitem{Hartnoll1} S. A. Hartnoll, C. P. Herzog and G. T. Horowitz, Building an AdS/CFT superconductor,
Phys. Rev. Lett. 101 (2008) 031601 [arXiv:0803.3295]


\bibitem{Hartnoll2} S. A. Hartnoll, C.P. Herzog and G. T. Horowitz, {\it JHEP} {\bf 12} 015
(2008) [arXiv:0810.1563]



\bibitem{Wu} J. P. Wu, Y. Cao, X. M. Kuang, and W. J. Li, Phys. Lett. B 697, 153 (2011)

 
\bibitem{Ma} D. Z. Ma, Y. Cao, and J. P. Wu, Phys. Lett. B 704, 604 (2011)
 \bibitem{Momeni} D. Momeni, N. Majd, and R. Myrzakulov, Europhys. Lett. 97, 61001 (2012)
\bibitem{Roychowdhury} D. Roychowdhury, Phys. Rev. D 86, 106009 (2012)
\bibitem{Momeni2} D. Momeni, M. R. Setare, and R. Myrzakulov, Int. J. Mod.Phys. A 27, 1250128 (2012)

\bibitem{Zhao} Z. Zhao, Q. Pan, and J. Jing, Phys. Lett. B 719, 440, (2013)




\bibitem{Myers} R.C. Myers, S. Sachdev, and A. Singh, Phys. Rev. D 83, 066017 (2011)



\bibitem{Zhao2} Z. Zhao, Q. Pan, S. Chen and J. Jing, Chinese Phys. Lett. 30, 121101 (2013)

\bibitem{Cai} R.G. Cai and D.W. Pang, Phys. Rev. D 84, 066004 (2011)


 



 

\bibitem{drummond}  I. T. Drummond and S. J. Hathrell, {\it Phys. Rev. D} {\bf 22} 343 (1980)

\bibitem{dereli1} T. Dereli and G. \"{U}\c{c}oluk, {\it Class. Q. Grav.} {\bf 7} 1109 (1990)

\bibitem{lambiase} G. Lambiase, S. Mohanty and G. Scarpetta,   {\it  JCAP} {\bf  07} 019 (2008)

\bibitem{bamba2}  K. Bamba,  S. Nojiri and S. D. Odintsov, {\it JCAP} {\bf 10} 045
(2008) [arXiv:0807.2575]

\bibitem{Kunze} K. E. Kunze {\it  Phys. Rev. D } {\bf  81}  043526
(2010) [arXiv:0911.1101]

\bibitem{dereli2} T. Dereli and \"{O}. Sert, {\it Phys. Rev. D} {\bf 83} 065005
(2011) [arXiv:1101.1177]

\bibitem{dereli4} T. Dereli and \"{O}. Sert,   { \it Mod. Phys. Lett. A}  {\bf  26} 1487
(2011) [arXiv:1105.4579]


\bibitem{sert2} \"{O}. Sert, M. Adak, An anisotropic cosmological solution to the Maxwell-Y(R) gravity   arXiv:1203.1531 [gr-qc]



\bibitem{Sert} \"{O}. Sert, M. Adak, Mod. Phys. Lett. A 28, 40,  1350190, (2013)




 \bibitem{gubser} S. S. Gubser, Phys. Rev. D 78, 065034 (2008)



\end{thebibliography}
\end{document}